\documentclass[]{spie}  

\usepackage{amsmath,amsfonts,amssymb}
\usepackage{graphicx}
\usepackage[colorlinks=true, allcolors=blue]{hyperref}
\usepackage{float}
\usepackage{subcaption}
\usepackage{hyperref}

\renewcommand{\vec}[1]{\mathbf{#1}}

\usepackage{setspace}

\title{Progressively-Growing AmbientGANs for learning Stochastic Object Models from imaging measurements}

\author[a]{Weimin Zhou}
\author[b]{Sayantan Bhadra}
\author[c]{Frank J. Brooks}
\author[c]{Hua Li}
\author[c]{Mark A. Anastasio}
\affil[a]{Department of Electrical and Systems Engineering,
\break Washington University in St$.\ $Louis, St$.\ $Louis, MO 63130, USA}
\affil[b]{Department of Computer Science and Engineering,
\break Washington University in St$.\ $Louis, St$.\ $Louis, MO 63130, USA}
\affil[c]{Department of Bioengineering, 
\break University of Illinois at Urbana-Champaign, Urbana, IL 61801, USA}

\authorinfo{Further author information: (Send correspondence to Mark A. Anastasio)\\Mark A. Anastasio: E-mail: maa@illinois.edu, Telephone: 1 217 333 1867}

\begin{document} 
\maketitle

\begin{abstract}
The objective optimization of medical imaging systems requires full characterization of all sources of randomness in the measured data, which includes
the variability within the ensemble of objects to-be-imaged.
This can be accomplished by establishing 
a stochastic object model (SOM) that describes the variability in the class of objects to-be-imaged.
Generative adversarial networks (GANs) can be potentially useful to establish SOMs because they hold great promise to learn generative models that describe the variability within an ensemble of training data.
However, because medical imaging systems record imaging measurements that are noisy and indirect representations of object properties,
GANs cannot be directly applied to establish stochastic models of objects to-be-imaged.
To address this issue,
an augmented GAN architecture named AmbientGAN was developed to establish SOMs from noisy and indirect measurement data.
 However, because the adversarial training can be unstable, the applicability of the
 AmbientGAN can be potentially limited. 
 In this work, we propose a novel training strategy---Progressive Growing of AmbientGANs (ProAGAN)---to stabilize 
 the training of AmbientGANs for establishing SOMs from noisy and indirect imaging measurements.
An idealized magnetic resonance (MR) imaging system and clinical MR brain images are considered.
The proposed methodology is evaluated by comparing signal detection performance computed by use of ProAGAN-generated synthetic images and images that depict the true object properties.
\end{abstract}

\keywords{stochastic object model, generative adversarial networks, signal detection, objective assessment of image quality}

\section{INTRODUCTION}
\label{sec:intro}  
It has been widely accepted that the optimization of medical imaging systems should be guided by objective measures of image quality (IQ) that quantify the performance of an observer at specific tasks\cite{barrett2013foundations, zhou2018learning, zhou2019learningIO, zhou2019learningHO, zhou2019approximating}.
The performance of a numerical observer  ideally should account for all sources of randomness in the measured data, which includes the variability within a group of objects being imaged\cite{barrett2013foundations, kupinski2003experimental, dolly2018learning, zhou2018learning, zhou2019learningIO, zhou2019learningHO, zhou2019approximating}. To achieve this, 
one can establish a stochastic object model (SOM) that can sample numerous realizations of objects from the ensemble of objects to-be-imaged. 
To establish a SOM that describes the to-be-imaged distribution, it is desirable to establish it from experimental measurements. Moreover, the established SOM should be independent of the imaging system.
Kupinski \emph{et al.} developed a method to fit SOMs by use of noisy imaging measurements\cite{kupinski2003experimental}.
However, the reported applications have been limited to lumpy and clustered lumpy object models\cite{kupinski2003experimental}.
In order to establish more complicated SOMs, Zhou \emph{et al.} implemented a deep learning method using an augmented generative adversarial network (GAN) architecture called an AmbientGAN\cite{bora2018ambientgan} to learn SOMs from noisy and indirect measurements\cite{zhou2019learning}. In that proof-of-concept study, a simple lumpy object model was considered\cite{zhou2019learning}. However, it is well-known that training of GANs can be unstable and, therefore, in practice, it can be difficult to learn the statistical properties of the to-be-imaged distribution.

A recently developed GAN training methodology---Progressive Growing of GANs (ProGANs)---holds great promise to stabilize the training of GANs and has been successfully employed to generate
high-quality synthetic images with high resolutions\cite{karras2017progressive}. 
Unlike the conventional training of GANs, in which all scales of the image distribution are learned simultaneously, 
the training of ProGANs starts with low-resolution images. Subsequently, network layers are added progressively to the generator and discriminator in synchrony to increase the image resolution for learning finer scale details of the image distribution.
Here, the generator and the discriminator have mirrored architectures. 
It has been shown that GAN training stability and the quality of the produced synthetic images can be improved significantly by use of this training strategy\cite{karras2017progressive}. However, because medical imaging systems record noisy and indirect measurements of object properties, ProGANs cannot be immediately
applied to establish stochastic models of objects to-be-imaged.

In this study, inspired by the ProGAN\cite{karras2017progressive} and the AmbientGAN\cite{bora2018ambientgan}, we propose a novel training methodology---Progressive Growing of AmbientGANs (ProAGANs)---to stably train AmbientGANs to establish SOMs from noisy and indirect imaging measurements. 
An idealized magnetic resonance (MR) imaging system and clinical MR brain images from NYU fastMRI Initiative database\cite{zbontar2018fastmri} were considered.
The proposed methodology was evaluated by comparing signal detection performance computed by use of ProAGAN-generated synthetic images and images that depict the true object properties.
\section{Background}
Consider a linear  discrete-to-discrete imaging system:
\begin{equation}
\vec{g} = \mathcal{H}\vec{f} + \vec{n},
\end{equation}
where $\vec{f}\in \mathbb{R}^{n^2}$ denotes a vector of an object image having dimension of $n\times n$, $\mathcal{H}: \mathbb{R}^{n^2} \rightarrow  \mathbb{R}^{m}$ denotes an imaging operator, $\vec{g}\in \mathbb{R}^{m}$ denotes a vector of imaging measurement data, and  $\vec{n}\in \mathbb{R}^{m}$ denotes measurement noise.

\subsection{GANs and AmbientGANs} 
Generative adversarial neural networks (GANs) were developed to learn data distributions by training a generator through an adversarial process with a discriminator\cite{goodfellow2014generative}.
Here, the generator is represented by a deep neural network having a set of weight parameters $\theta_G$ and a mapping function $G(\cdot\ ; \theta_G)$: $\mathbb{R}^k \rightarrow \mathbb{R}^{n^2}$ that maps a latent vector $\vec{z}\in \mathbb{R}^k$ to a synthetic image $\hat{\vec{f}} = G(\vec{z} ; \theta_G)$. The discriminator is represented by another deep neural network having a set of weight parameters $\theta_D$ and a mapping function $D(\cdot\ ; \theta_D)$: $\mathbb{R}^{n^2} \rightarrow \mathbb{R}$ that maps a real or synthetic image to a real-valued scalar $s$ that can be used to distinguish between real and synthetic images. However, because medical imaging systems acquire noisy and indirect imaging measurements $\vec{g}$, GANs cannot be directly employed to establish SOMs.

An augmented GAN architecture named AmbientGAN was developed to learn SOMs from noisy and indirect measurements of object properties\cite{bora2018ambientgan,zhou2019learning}.
Similar to GANs, AmbientGANs comprise a generator and a discriminator. 
The generator is trained to produce synthetic digital images that depict the object properties. Synthetic imaging measurements are subsequently simulated by applying the corresponding imaging operator $\mathcal{H}$ to the finite-dimensional representations of the object property produced by the generator.
The discriminator that maps $\mathbb{R}^{m} \rightarrow \mathbb{R}$ is applied to distinguish between real and synthetic imaging measurements. 
The training process of AmbientGANs can be represented by a two-player minimax game:
\begin{equation} \label{eq:AGAN}
\min_{\theta_G} \max_{\theta_D} {E_{\vec{g}\sim p_\vec{g}}} [l\left(D(\vec{g}; \theta_D)\right)] + {E_{\hat{\vec{g}} \sim p_{\hat{\vec{g}}}}} [l(1- D\left( \hat{\vec{g}}; \theta_D \right) )].
\end{equation} 
Here, $\hat{\vec{g}} = \mathcal{H}\hat{\vec{f}}+ \vec{n}$, where $\hat{\vec{f}} = G(\vec{z}; \theta_G)$, and $l(\cdot)$ represents an objective function.  
It has been shown that when $p_\vec{f}$ uniquely induces $p_\vec{g}$, and both the discriminator and the generator possess sufficient capacity, $p_{\hat{\vec{f}}}$ approximates $p_{\vec{f}}$
when the global optimal of the minimax game is achieved\cite{goodfellow2014generative, bora2018ambientgan}. 

\subsection{Progressive Growing of GANs} 
Progressive Growing of GANs (ProGANs) were developed to stabilize the training of GANs\cite{karras2017progressive}. 
Unlike the conventional GANs, in which the generator and the discriminator are trained to learn all scales of image distribution by use of full resolution images  through the whole training process,
the training of ProGANs start with training the first few layers of the generator and discriminator on low-resolution images that are down-scaled from the original images.
Higher-resolution images are subsequently employed and more layers of the generator and discriminator are progressively included in the training to learn finer scale details of the image distribution.
It has been shown that the ProGAN training strategy results in significantly improved training stability and produces synthetic images with the state-of-the-art image quality\cite{karras2017progressive}.
However, like other GANs, ProGANs cannot be immediately applied to learn SOMs from noisy and/or indirect imaging measurements.

\section{Progressive Growing of Ambient-GANs} 
\begin{figure}[H]
\centering
 \includegraphics[width=0.9\linewidth]{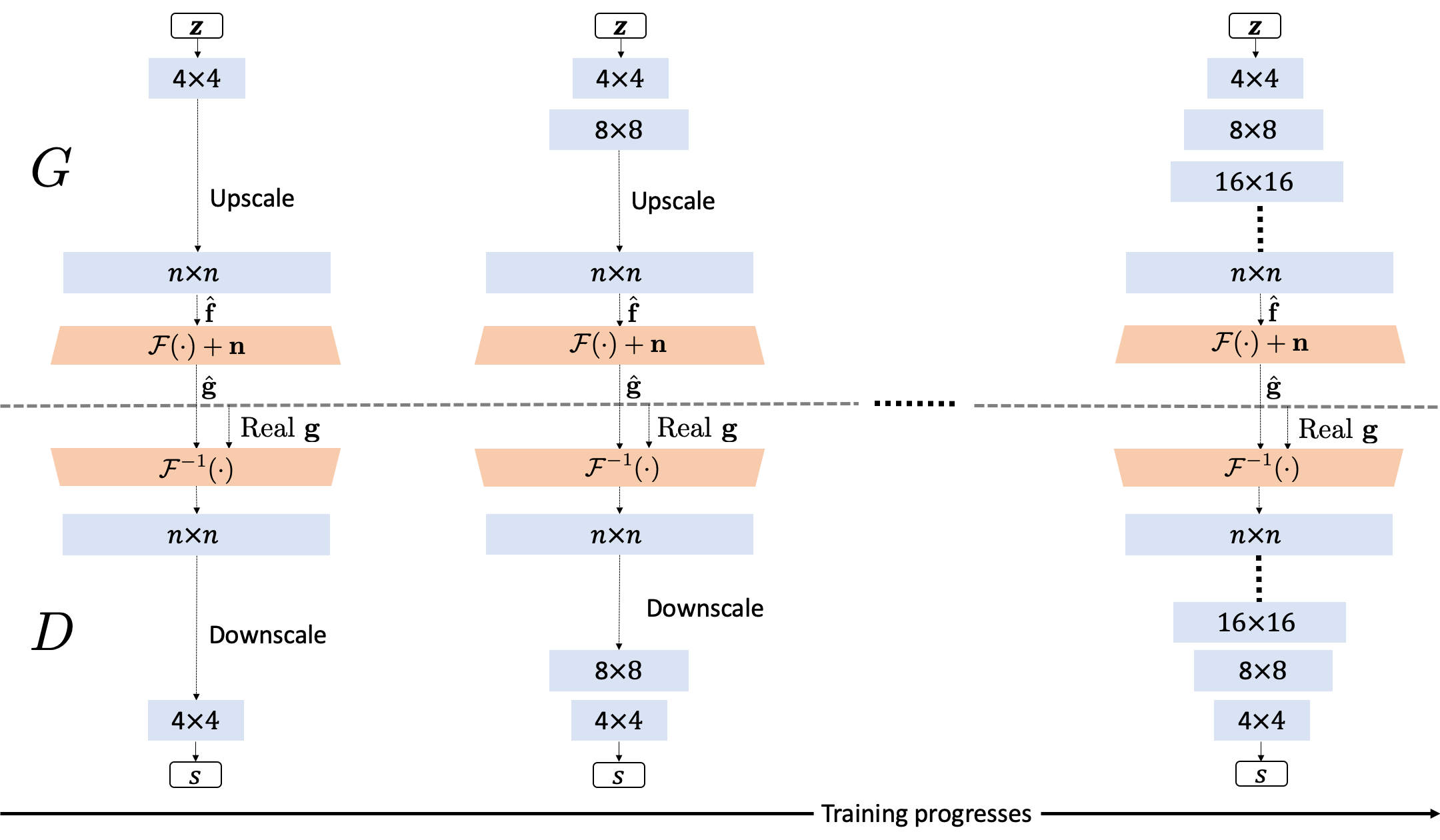}
\caption{ProAGAN training for the considered MR imaging system. The training started with low-resolution images, and then increased the image resolution progressively by including more layers to the generator and the discriminator in the training. The synthetic k-space measurements were simulated by use of the DFT. The IDFT was applied to reconstruct real and synthetic object images. The discriminator was trained to distinguish between real and synthetic reconstructed images.}
  \label{fig:arc}
 \end{figure}

In this work, we propose a novel training strategy named Progressive Growing of AmbientGANs (ProAGANs) to stably train the AmbientGANs for establishing SOMs from noisy and indirect imaging measurements.
A MR imaging system that fully samples k-space data was considered: $\vec{g} = \mathcal{F}(\vec{f}) + \vec{n}$, where $\mathcal{F}(\cdot)$ denotes a 2D discrete Fourier transform (DFT).
The generator in the proposed ProAGAN was trained to synthesize images that depict object properties in the to-be-imaged distribution.
However, because the MR imaging system records noisy k-space measurement data, the discriminator cannot be applied directly to distinguish between real and synthetic object images. 
In our proposed architecture of ProAGAN, a 2D inverse discrete Fourier transform (IDFT) $\mathcal{F}^{-1}$ is included in the training of ProAGAN to reconstruct object images from k-space measurements, and the discriminator is trained 
to distinguish between real and synthetic reconstructed images. 
The goal is to learn the distribution of images that depict to-be-imaged object properties by progressively training a generator by competing against a discriminator that distinguishes between real and synthetic reconstructed images.
This training process is illustrated in Fig. \ref{fig:arc}.

\section{Numerical studies}
\begin{figure}[H]
\centering
\begin{subfigure}[b]{0.24\textwidth}
   \centering
 \includegraphics[width=1.0\linewidth]{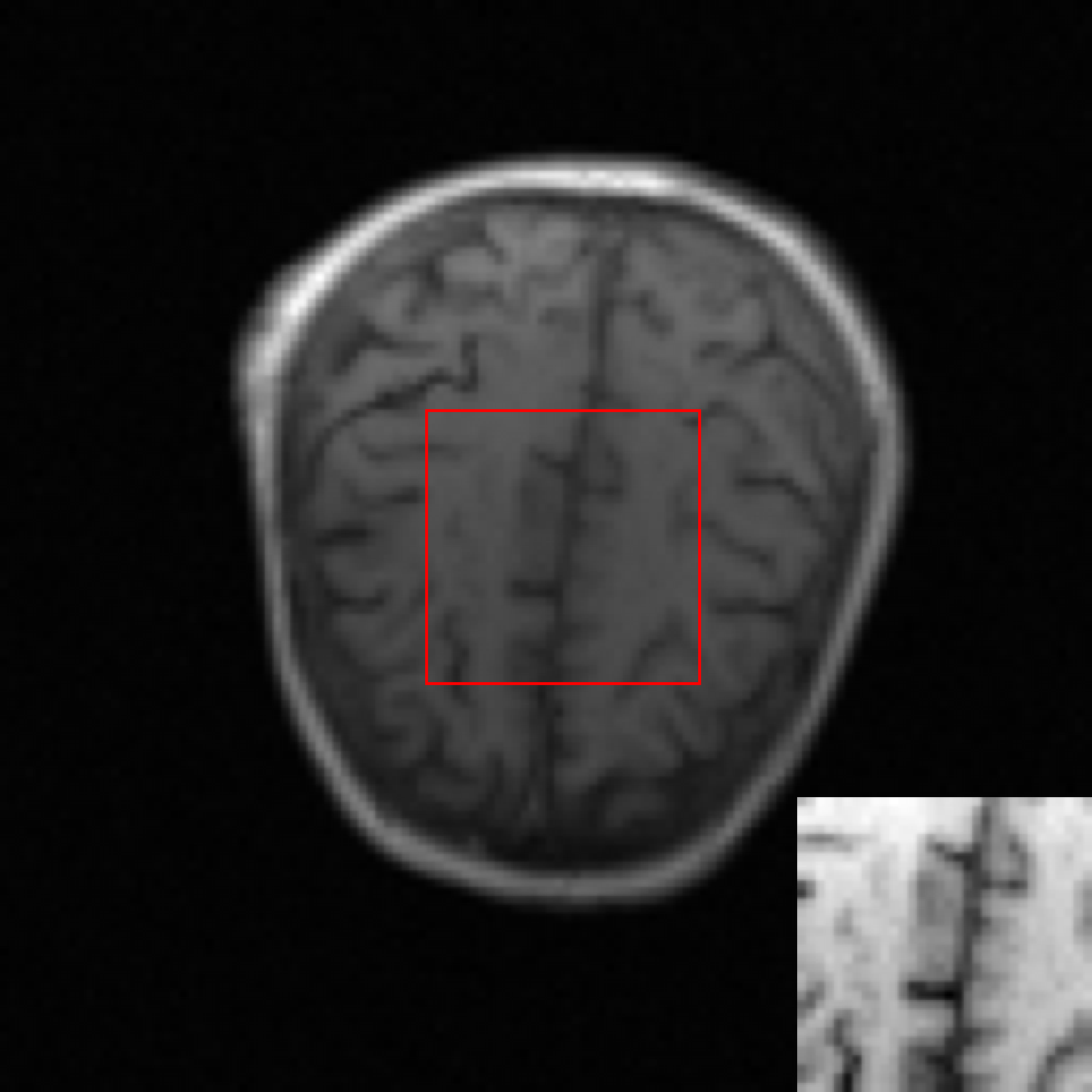}
 \caption{}
 \end{subfigure}
 \begin{subfigure}[b]{0.24\textwidth}
  \centering
 \includegraphics[width=1.0\linewidth]{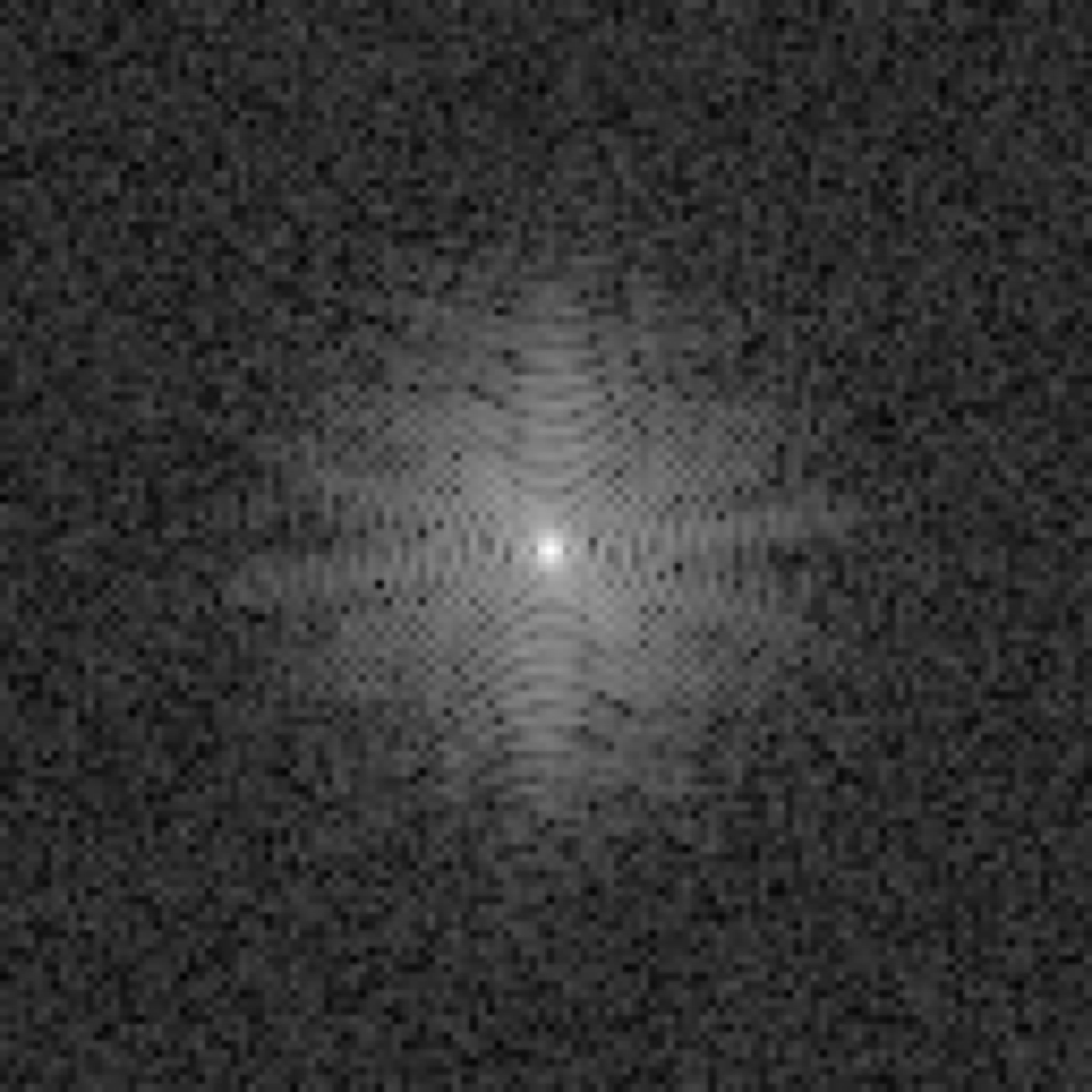}
 \caption{}
 \end{subfigure}
  \begin{subfigure}[b]{0.24\textwidth}
  \centering
 \includegraphics[width=1.0\linewidth]{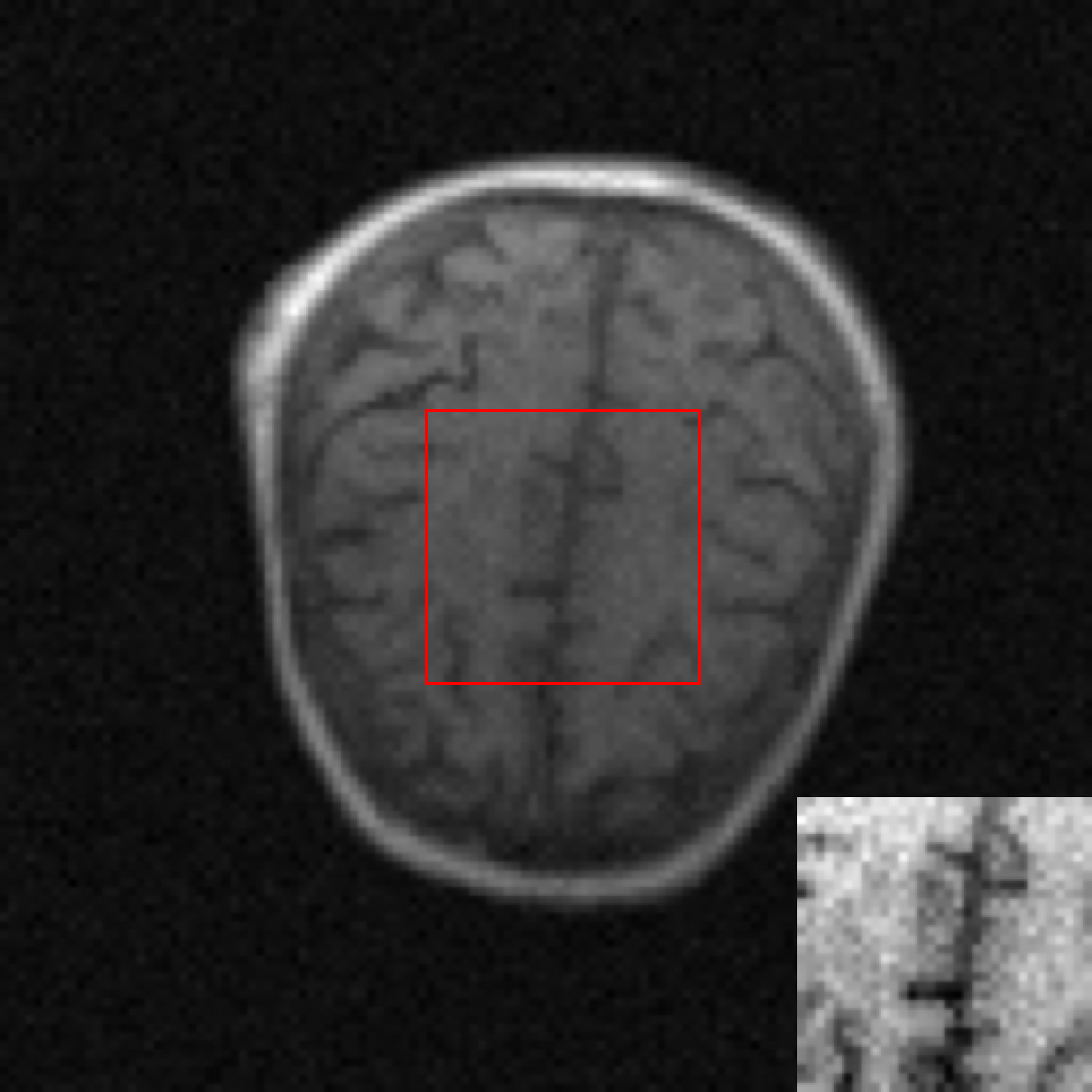}
 \caption{}
 \end{subfigure}
 \vspace{0.2cm}
 \caption{(a) A MR brain image. (b) k-space measurement data corresponding to (a). (c) The IDFT of (b). A central region in (a) and (c) was displayed to better visualize the noise in the reconstructed image.}
 \label{fig:obj}
\end{figure}

A simulation study was conducted.
MR brain images in the NYU fastMRI Initiative database \cite{zbontar2018fastmri}  (\url{https://fastmri.med.nyu.edu/}) were employed to form an ensemble of images that depict object properties sampled from the unknown SOM. Specifically,
3000 T1 weighted brain MR images corresponding to the magnetic field strength of 3T were selected, and these 3000 images were 
resized to the dimension of $128\times 128$ to be employed as real object images.
Fully-sampled MR k-space data of these 3000 object images were simulated, and complex Gaussian noise were added to the k-space data.
These 3000 noisy k-space measurement data formed the training dataset. An example of MR brain images, its corresponding k-space measurement data and the reconstructed image (i.e., IDFT of k-space measurement data) are shown in Fig. \ref{fig:obj}.
 
 \subsection{Training details}
The proposed ProAGAN was implemented in Tensorflow\cite{abadi2016tensorflow} by modifying the ProGAN codes to apply the architecture illustrated in Fig. \ref{fig:arc}. The codes for the ProGAN can be found at \url{https://github.com/tkarras/progressive_growing_of_gans}. The ProAGAN was trained by use of 4 NVIDIA Tesla V100 GPUs. The Adam algorithm \cite{kingma2014adam}, which is a stochastic gradient algorithm, was employed to train ProAGANs.

 \subsection{Hotelling observer validation studies}
\begin{figure}[H]
   \centering
 \includegraphics[width=0.77\linewidth]{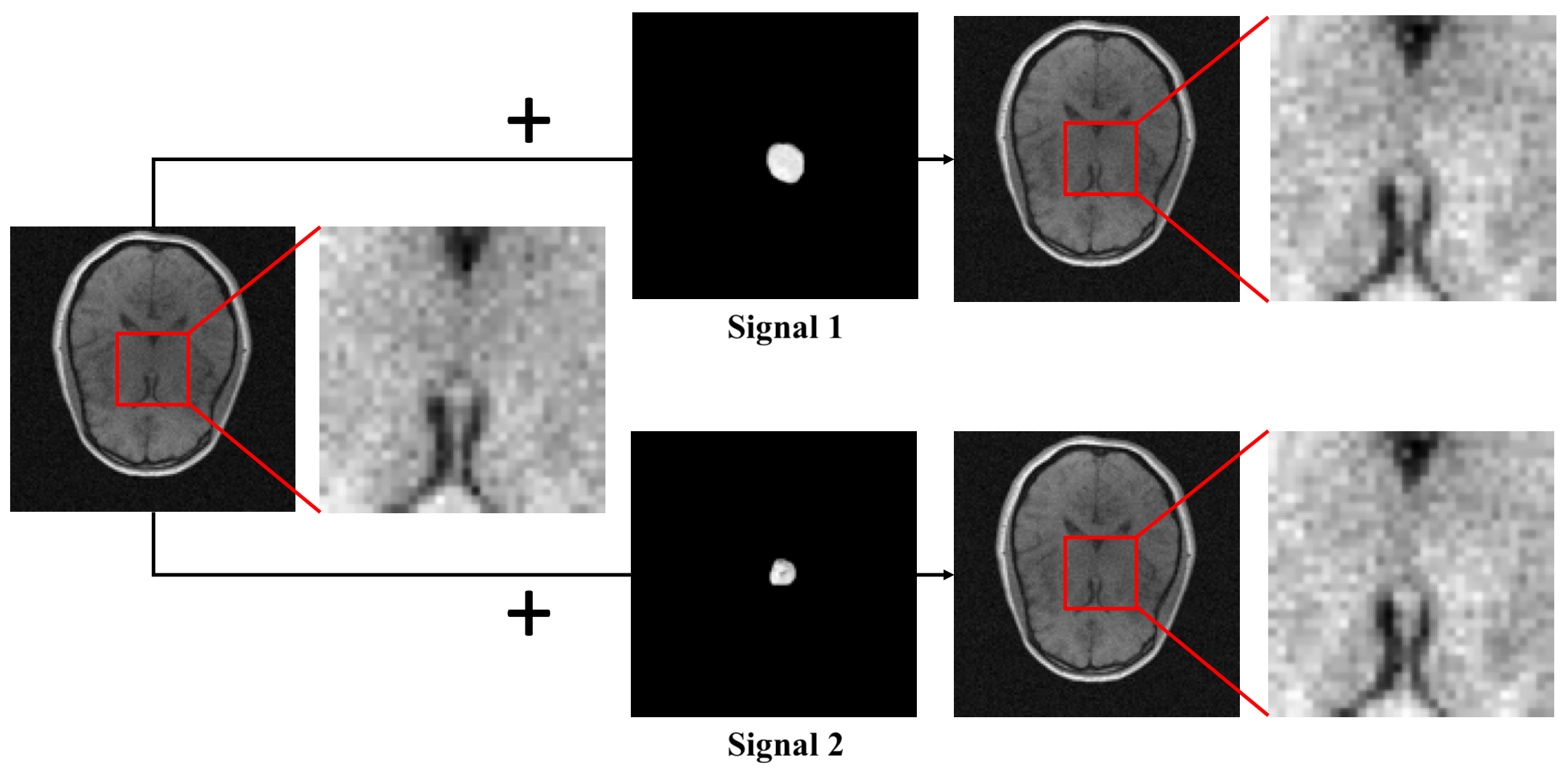}
 \caption{An example of signal-absent images, two signals, and the corresponding signal-present images. The region of interest (ROI) where the signals may be presented is displayed.}
 \label{fig:signal_imgs}
\end{figure}

Signal-known-exactly (SKE) binary signal detection tasks were conducted to assess the learned SOM in a task-dependent way. 
The considered signal detection tasks require an observer to classify a MR brain image as satisfying signal-absent hypothesis ($H_0$) or signal-present hypothesis ($H_1$):
\begin{subequations}
\label{eq:imgH_s}
\begin{align}
H_{0}:&\  \mathbf{g} = \mathbf{f} + \mathbf{n}, \\
H_{1}:&\  \mathbf{g} = \mathbf{f} + \mathbf{s} + \mathbf{n},
\end{align}
\end{subequations}
where $\mathbf{s}$ denotes the signal to be detected, and $\mathbf{n}$ is independent and identically distributed Gaussian noise.
Two different signals represented by two artificial tumors were considered. These signals are shown in Fig. \ref{fig:signal_imgs}.

The signal detection tasks were performed on a region of interest (ROI) that was centrally located in the images. The ROIs had dimension of $32\times 32$.
The covariance matrices corresponding to the real and ProAGAN-produced images were computed by use of 3000 ROIs extracted from the real and ProAGAN-produced images, respectively.
These covariance matrices were subsequently employed to computed the Hotelling observer (HO) by use of a covariance matrix decomposition\cite{barrett2013foundations}.
The  performances of the HOs were evaluated on a testing dataset that comprised 500 pairs of
signal-absent and signal-present images. The receiver operating characteristic (ROC) curves and AUC values corresponding to the ProAGAN-produced images were
compared to those corresponding to the real images. The ‘proper’ binormal model\cite{pesce2007reliable} was employed to fit the ROC curves.
\section{Results}
\begin{figure}[H]
\centering
\begin{subfigure}[b]{0.19\textwidth}
   \centering
 \includegraphics[width=1.0\linewidth]{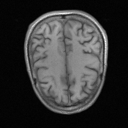}
 \caption{}
 \end{subfigure}
 \begin{subfigure}[b]{0.19\textwidth}
  \centering
 \includegraphics[width=1.0\linewidth]{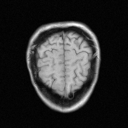}
 \caption{}
 \end{subfigure}
  \begin{subfigure}[b]{0.19\textwidth}
  \centering
 \includegraphics[width=1.0\linewidth]{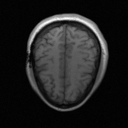}
 \caption{}
 \end{subfigure}
  \begin{subfigure}[b]{0.19\textwidth}
  \centering
 \includegraphics[width=1.0\linewidth]{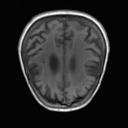}
 \caption{}
 \end{subfigure}
   \begin{subfigure}[b]{0.19\textwidth}
  \centering
 \includegraphics[width=1.0\linewidth]{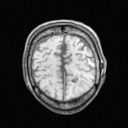}
 \caption{}
 \end{subfigure}

\vspace{0.3cm}
  \begin{subfigure}[b]{0.19\textwidth}
   \centering
 \includegraphics[width=1.0\linewidth]{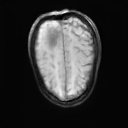}
 \caption{}
 \end{subfigure}
 \begin{subfigure}[b]{0.19\textwidth}
  \centering
 \includegraphics[width=1.0\linewidth]{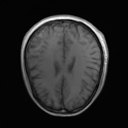}
 \caption{}
 \end{subfigure}
  \begin{subfigure}[b]{0.19\textwidth}
  \centering
 \includegraphics[width=1.0\linewidth]{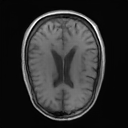}
 \caption{}
 \end{subfigure}
  \begin{subfigure}[b]{0.19\textwidth}
  \centering
 \includegraphics[width=1.0\linewidth]{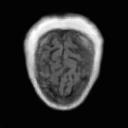}
 \caption{}
 \end{subfigure}
   \begin{subfigure}[b]{0.19\textwidth}
  \centering
 \includegraphics[width=1.0\linewidth]{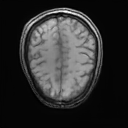}
 \caption{}
 \end{subfigure}
 \vspace{0.1cm}
 \caption{(a)-(e) Real MR brain images.  (f)-(j) Synthetic MR brain images produced by the ProAGAN.}
 \label{fig:img_real_fake}
\end{figure}

Examples of real and synthetic images of object properties produced by the ProAGAN are shown in Fig. \ref{fig:img_real_fake}. 
The synthetic images are promising in terms of visually mimicking the real images.

\begin{figure}[H]
\centering
\begin{subfigure}[b]{0.16\textwidth}
   \centering
 \includegraphics[width=1.0\linewidth]{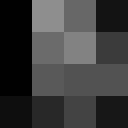}
 \caption{}
 \end{subfigure}
 \begin{subfigure}[b]{0.16\textwidth}
  \centering
 \includegraphics[width=1.0\linewidth]{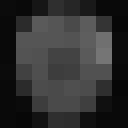}
 \caption{}
 \end{subfigure}
  \begin{subfigure}[b]{0.16\textwidth}
  \centering
 \includegraphics[width=1.0\linewidth]{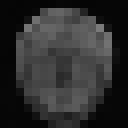}
 \caption{}
 \end{subfigure}
  \begin{subfigure}[b]{0.16\textwidth}
  \centering
 \includegraphics[width=1.0\linewidth]{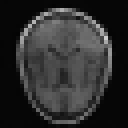}
 \caption{}
 \end{subfigure}
   \begin{subfigure}[b]{0.16\textwidth}
  \centering
 \includegraphics[width=1.0\linewidth]{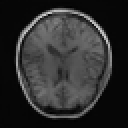}
 \caption{}
 \end{subfigure}
    \begin{subfigure}[b]{0.16\textwidth}
  \centering
 \includegraphics[width=1.0\linewidth]{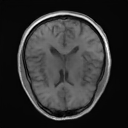}
 \caption{}
 \end{subfigure}
  \caption{(a)-(f) Synthetic images produced by the generator at different training steps corresponding to the resolution of $4\times 4$, $8\times 8$, $16\times 16$, $32\times 32$, $64\times 64$, and $128\times 128$.}
 \label{fig:img_grow}
\end{figure}

To illustrate the progressive growing of ProAGAN, synthetic images at different training steps with different resolution levels are shown in Fig. \ref{fig:img_grow}.
It is demonstrated that the generator was progressively established in a stable way.

\begin{figure}[H]
\centering
\hspace{1cm} \includegraphics[width=0.9\linewidth]{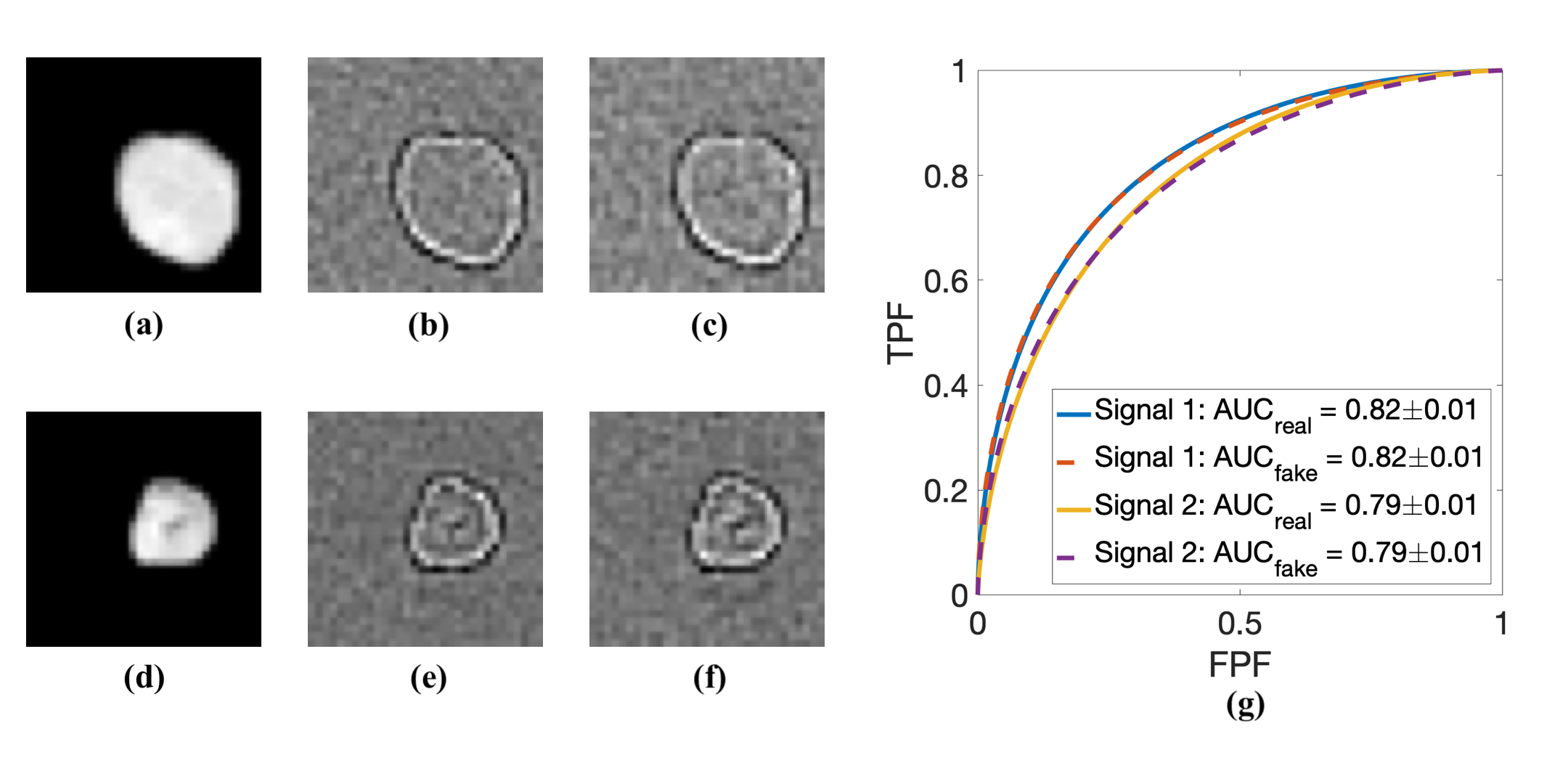}
\caption{(a) ROI corresponding to signal 1. (b) Hotelling template corresponding to signal 1 that was computed by use of ROIs of real images.  (c) Hotelling template corresponding to signal 1 that was computed by use of ROIs of synthetic images produced by the ProAGAN. (d) ROI corresponding to signal 2.  (e) Hotelling template corresponding to signal 2 that was computed by use of ROIs of real images. (f) Hotelling template corresponding to signal 2 that was computed by use of ROIs of synthetic images produced by the ProAGAN. (g) ROC curves and AUC values corresponding to real and synthetic images. The ROC curves almost overlap.}
  \label{fig:HO}
 \end{figure}

The Hotelling templates that were computed by use of the real images and the ProAGAN-produced images are compared in Fig. \ref{fig:HO}.
The ROC curves and AUC values corresponding the real images and the ProAGAN-produced images are compared in Fig. \ref{fig:HO} (g), in which the solid curves and the dashed curves correspond to the HOs computed by use of real images and synthetic images produced by the ProAGAN, respectively. The ROC curves are almost identical.

\section{Conclusion}
This study provides a novel training methodology---Progressive Growing of AmbientGANs---to stably train AmbientGANs for learning SOMs from noisy and indirect measurement data. In this preliminary study, an idealized magnetic resonance (MR) imaging system was considered, and a task-based validation study was conducted by use of the Hotelling observer. It is demonstrated that the proposed method is able to establish a SOM by use of noisy k-space measurement data. From the perspective of the HO, for the signal detection tasks considered, it was found that the images synthesized by the learned SOM
contained nearly the same task-specific information as the true object property images.

\section*{ACKNOWLEDGMENT}       
This research was supported in part by NIH awards EB020604, EB023045, NS102213, EB028652, and NSF award DMS1614305.

\bibliography{PAmbientGAN.bib}
\bibliographystyle{spiebib} 

\end{document}